\documentclass[superscriptaddress]{revtex4}
\usepackage{amsmath,amssymb}
\usepackage{bm,graphics,mathtools}
\usepackage{color}
\usepackage[letterpaper, margin=1.25in]{geometry}
\begin{document}
\title{Characterizing Localized Surface Plasmons using Electron Energy-Loss Spectroscopy}
\author{Charles Cherqui}
\affiliation{Department of Chemistry, University of Washington, Seattle, WA 98195}
\author{Niket Thakkar}
\affiliation{Department of Applied Mathematics, University of Washington, Seattle, WA 98195}
\author{Guoliang Li}
\affiliation{Department of Chemistry and Biochemistry, University of Notre Dame, Notre Dame, IN 46556}
\author{Jon P. Camden}
\affiliation{Department of Chemistry and Biochemistry, University of Notre Dame, Notre Dame, IN 46556}
\author{David Masiello}
\affiliation{Department of Chemistry, University of Washington, Seattle, WA 98195}
\affiliation{Department of Applied Mathematics, University of Washington, Seattle, WA 98195}
%\documentclass[amsmath,amssymb]{revtex4-1}
%\usepackage{bm,graphics,mathtools}
%\usepackage{color}
%\jname{Xxxx. Xxx. Xxx. Xxx.}
%\jvol{00}
%\jyear{2015}
%\doi{10.1146/((please add article doi))}
%
%% Document starts
%\begin{document}
%
%% Title
%\title{Characterizing Localized Surface Plasmons using Electron Energy-Loss Spectroscopy}
%% Author/affiliation
%\author{Charles Cherqui,$^1$ Niket Thakkar,$^2$ Guoliang Li,$^3$ Jon P. Camden$^3$,  David J. Masiello$^{1,2}$
%\affil{$^1$Department of Chemistry, University of Washington, Seattle, WA 98195
%email: masiello@chem.washington.edu} 
%\affil{$^2$Department of Applied Mathematics, University of Washington, Seattle, WA 98195}
%\affil{$^3$Department of Chemistry and Biochemistry, University of Notre Dame, Notre Dame, IN 46556; email: jon.camden@nd.edu}}

%%%%%%%%%%%%%%%%%%%%%%%%%%%%%%%%%%%%%%%%%%%%%%%%%%%%%%%%%%%%%%%%%%%%%%%%%

\setlength{\pdfpageheight}{11in}
\setlength{\pdfpagewidth}{8.5in}
\hoffset = 0.03125truein
\voffset = -0.03125truein
%%%%%%%%%%%%%%%%%%%%%%%%%%%%%%%%%%%%%%%%%%%%%%%%%%%%%%%%%%%%%%%%%%%%%%%%%
\linespread{1.25}
\selectfont

% Abstract
\begin{abstract}
Electron energy-loss spectroscopy (EELS) offers a window to view nanoscale properties and processes. When performed in a scanning transmission electron microscope, EELS can simultaneously render images of nanoscale objects with sub-nanometer spatial resolution and correlate them with spectroscopic information of $\sim10 - 100$ meV spectral resolution. Consequently, EELS is a near-perfect tool for understanding the optical and electronic properties of individual and few-particle plasmonic metal nanoparticles assemblies, which are significant in a wide range of fields. This review presents an overview of basic plasmonics and EELS theory and highlights several recent noteworthy experiments involving the electron-beam interrogation of plasmonic metal nanoparticle systems.
\end{abstract}

% Keywords
%\begin{keywords}
%electron energy-loss spectroscopy, localized surface plasmons, STEM/EELS
%\end{keywords}

\maketitle

% to generate article TOC
%\tableofcontents

% Head 1
\section{Introduction}
At optical frequencies metals screen applied fields imperfectly, allowing light to travel roughly tens of nanometers into metallic structures \cite{kreibig1985optical}. Thus, metal nanoparticles (MNPs), whose physical extent is on par with the penetration depth, differ from the bulk in that light can penetrate them entirely. From just this observation, it is evident that MNP conduction electrons can be uniformly forced by light and made to exhibit collective motion. Moreover, this motion must be strongly dependent on MNP geometry since the electrons are restricted by its boundary. Indeed, given enough time to respond to an applied field, the MNP's conduction electrons accelerate until they reach the particle's surface, leaving a positive ionic background on the other side and polarizing the particle. When the applied field is removed, electrons accelerate towards the exposed ionic background, overshooting their equilibrium position, collectively oscillating before decaying back to their initial state. This collective excitation is called a plasmon. In general, plasmons can be excited by light or by near-field sources; they exist in the bulk and on surfaces; the latter can be localized to nanoscopic particles or propagate on macroscopic ones. Those that propagate are known as surface plasmon polaritons (SPPs), while those localized to MNPs are called localized surface plasmons (LSPs). 

MNPs are capable of supporting a broad range of modes that can be grouped into two categories: bright and dark. The former have a net dipole moment and couple to light, while the latter do not. 
%requires that the spatial profile of the driving field has enough curvature to excite it.
% Alternatively, dark modes can be excited by light if they exist on a MNP large enough to sample the curvature of light over an optical cycle or by speed of light delays as information travels across the MNP, in which case they are no longer strictly speaking "dark" since they weakly couple to the radiation-field. 
Since light may be used to drive bright LSPs, so too can they radiate, thereby focusing far-field radiation to sub-diffraction-limited length scales while simultaneously acting as nanoscopic antennas. Bright LSPs play a pivotal role in a number of areas, including enhanced molecular-optical spectroscopy \cite{jeanmaire1977surface, Kneipp1997,Nie_single_molecules_1997}, chemical sensing and catalysis \cite{Malinsky_chemical_sensing_2001}, photothermal cancer therapy \cite{Huang_cancer_therapy_2006}, and solar-energy conversion \cite{Linic_efficient_conversion_2011}. In contrast, dark LSPs lose energy to their environments nonradiatively and therefore have fewer available decay pathways, leading to higher quality factors \cite{herzog2013dark,gomez2013dark}. Recently, research into dark modes has intensified as plasmonic applications call for LSPs with just these properties \cite{herzog2013dark,gomez2013dark}. Dark LSPs are well-suited for novel applications such as nanoscale heating \cite{Arash_nanoscale_heating_2015,baldwin14} or as tunable sources of energetic electrons \cite{Mukherjee_hot_electron_2013}. Taken together, both bright and dark LSPs offer a unique means to direct the flow of light, energy, and charge \cite{ozbay2006plasmonics} on the nanoscale, and as a result, the characterization of single nanoparticles and aggregates continues to play a central role in the development and advancement of nanoscience \cite{maier2007plasmonics}. 

Spectroscopic methods employing plane-wave excitation sources are commonly used to study LSPs. These include dark-field optical microscopy (DFOM) \cite{Mock_composite_plasmon_2002, Murray_Plasmonic_2007, Sonnichsen_nanorod_2005, Schultz_single_molecule_2000}, broadband extinction spectroscopy \cite {Slaughter_single_particle_2010, Lindfors_confocal_microscopy_2004}, photothermal imaging \cite{Chang_Plasmonic_nanorod_2010, Berciaud_Photothermal_2004} and non-linear confocal microscopy \cite{Ghenuche_nanoantennas_2008}. The spatial resolution of these techniques is limited by the diffraction limit of light ($\sim$200 nm), leaving the spatial variation of the nanoparticle's LSP modes obscured. Moreover, by their very nature, methods employing far-field radiation are incapable of interrogating dark LSPs. To improve upon these limitations, methods based on transforming far-field light into a near-field source, such as near-field scanning optical microscopy (NSOM) have successfully achieved spatial resolution below 10 nm \cite{Wessel_Surface_enhanced_1985}, but images and spectra can be difficult to interpret since the near-field tip significantly perturbs the local electric field of the MNP aggregate. For details, we refer the reader to the following review articles \cite{Duo_Recent_developments_2010, Olson_Optical_characterization_2015}. 

Electron energy-loss spectroscopy (EELS) offers an elegant solution to the limitations described above by using a relativistic electron beam in place of optical excitation sources. The basic concept behind EELS is to bombard a target with electrons and collect those that have been reflected or transmitted to determine the change in their kinetic energy. Then, by energy conservation, this change corresponds to the excitation energies of the target's modes. Unlike plane-wave probes, the electron beam provides a highly localized electric field with which to interrogate a MNP. This gives EELS access to both bright and dark LSPs while providing a near unlimited spectral range along with a spatial resolution limited only by the sub-{\AA}ngstrom de Broglie wavelength of a relativistic electron.

The first study on the kinetic energy of electrons scattered off of thin Cu films was preformed by Rudberg in 1930 \cite{rudberg1930characteristic}. He measured the electron intensity as a function of loss energy and showed that the spectrum could be correlated to the chemical composition of the sample. The loss spectra of transmitted electrons was first observed by Ruthemann in 1941 \cite{ruthemann1941diskrete}, who measured the loss energy of electrons passing though thin Al films. This work is of particular importance since the peaks recorded in the data were later interpreted as bulk plasmons by Bohm and Pines \cite{bohm1953collective,pines1952collective}, laying the groundwork for modern condensed-matter theory and providing a rigorous justification for the free electron gas approximation assumed in the Drude model of metals \cite{drude1900elektronentheorie}.  EELS also played a central role in the discovery of surface plasmons. In 1957, Ritchie predicted the existence of surface modes in the loss spectra of thin metal films. This was quickly confirmed experimentally by Powell and Swan \cite{powell1959origin} and shortly after by Stern and Ferrell \cite{stern1960surface}, who also named the quanta of these modes surface plasmons. 

The study of LSPs using EELS was first suggested theoretically in 1968 by Fujimoto and Komaki in Japan \cite{fujimoto1968plasma} and independently by Crowell and Ritchie in the United States \cite{crowell1968radiative}. While both predicted the existence of dark modes, the latter also predicted the radiative decay of the lowest order bright mode, which was confirmed shortly after in 1972 \cite{kokkinakis1972observation}. The conclusive observation of dark modes would have to wait for advances in high-resolution scanning transmission electron microscopy (STEM) \cite{Batson1980477, Batson1982277, Cowley1982587,PhysRevB.25.1401,Colliex1985131,JMI:JMI534}, which allowed for unprecedented control over the position of the electron beam together with unprecedented energy resolution. The inclusion of EELS in these devices provided conclusive evidence of the existence of dark modes in spherical MNPs in 1987 \cite{wang1987surface,wang1987excitation}.

% Margin note

%\begin{marginnote}
%\entry{STEM}{Scanning Transmission Electron Microscope}
%\entry{EELS}{Electron Energy-Loss Spectroscopy}
%\entry{LSP}{Localized Surface Plasmon}
%\entry{SPP}{Surface Plasmon Polariton}
%\entry{MNP}{Metal Nanoparticle}
%\end{marginnote}

% Margin note

\begin{figure}
\centering
\includegraphics[scale=1]{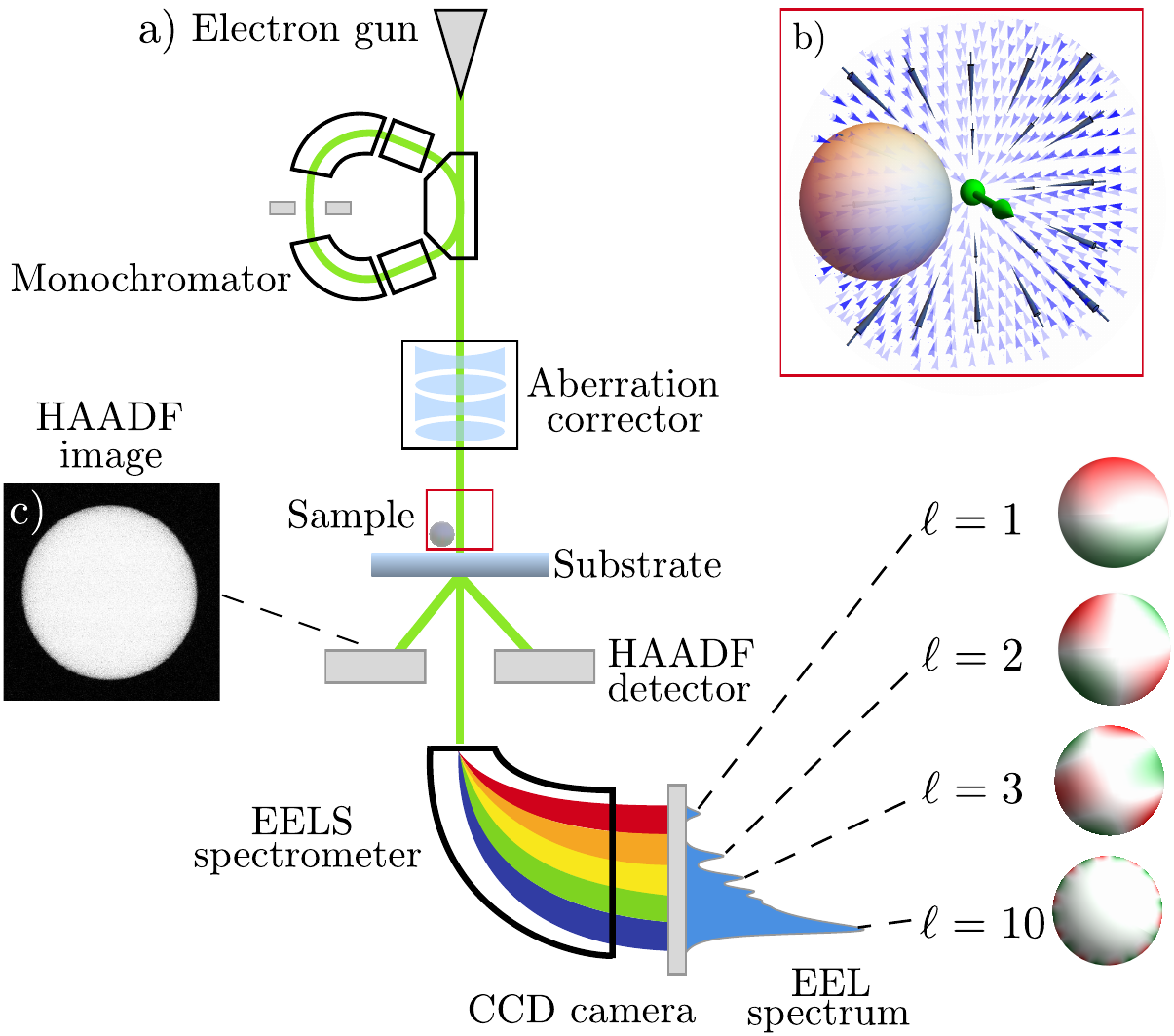}
\caption{(a) Schematic of an EELS experiment performed in a STEM equipped with a monochromator to narrow the energy distribution of electrons and an aberration corrector to minimize the electron probe size. After passing through or by the sample, electrons scattered by small angles go through a spectrometer. The energy dispersive plane is projected onto a CCD camera, yielding an EELS spectrum. (b) Schematic of a fast electron in close proximity to a nanoparticle surface. The electric field of an electron moving at $0.9c$, exhibiting a flattened electric field in the direction perpendicular to its trajectory. (c) Electrons may also scatter off target nuclei and are collected by a high-angle annular dark field (HAADF) detector, producing an image of the sample.}
\label{EELS Schematic}
\end{figure}

In the last 30 years the field of electron microscopy has seen remarkable gains in the spatial and energetic resolution of STEM/EELS experiments \cite{de2010optical,egerton2011electron}. {\bf Figure \ref{EELS Schematic}} illustrates how modern EELS experiments are performed in a STEM. First, electrons are accelerated to $\sim10-10^2$ keV. Most electrons pass by or through the target without interaction, however, some electrons are scattered to wide angles by the target's atomic nuclei and collected by a high-angle annular dark field (HAADF) detector to create a nanoscale image of the target. Beyond that, a small number of electrons inelastically scatter and impart a fraction of their kinetic energy into the target. The ratio of the number of inelastically scattered electrons to the total number of electrons defines the electron energy-loss probability. The spectral decomposition of the loss probability is called the loss-probability spectrum and can be interpreted as the probability to excite a target mode of a particular energy at a particular electron beam position.

\begin{figure}
\centering
\includegraphics[scale=0.44]{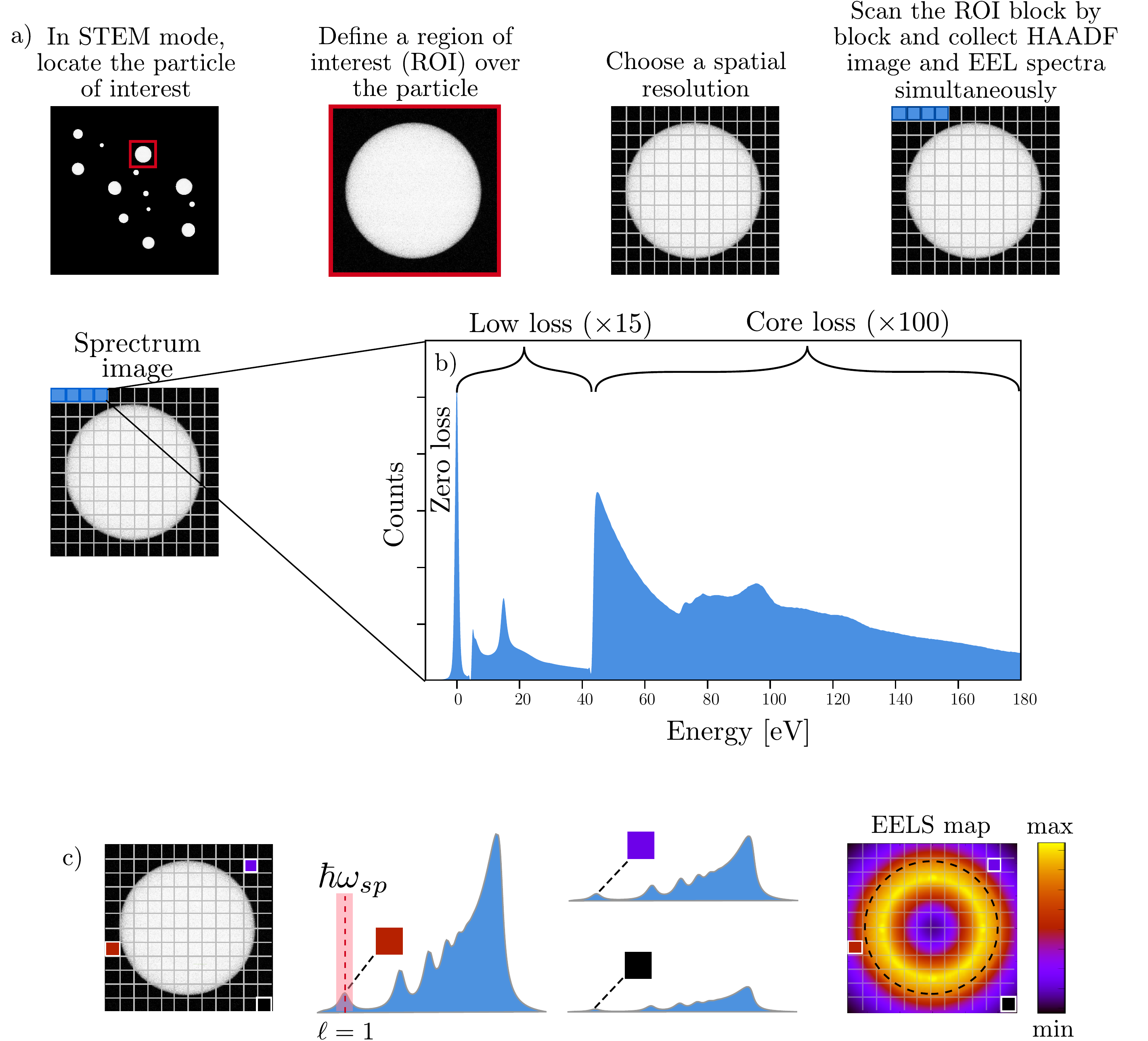}
\caption{(a) Schematic of the EELS spectrum image acquisition. A spectrum image is obtained using the HAADF detector to define a region of interest (ROI) over a nanoparticle. The ROI is then pixelated with a chosen spatial resolution and scanned pixel by pixel, simultaneously recording a HAADF image and EEL spectrum. (b) Typical experimental EEL spectrum exhibiting the zero loss peak, the low-loss region and the core-loss region. (c) Schematic of LSP mode mapping. The spectrum of each pixel is evaluated at the LSP resonance frequency $\hbar\omega_{sp}$. The height of the peak or the area under a fitted Lorentzian is then organized into a color map.}
\label{SpectraMaps}
\end{figure}

In spectrum image mode (SI mode), the beam may be rastered across a region of interest (ROI) that may include the target, and at each position both HAADF and energy-loss signals can be simultaneously acquired \cite{Kociak_Spatially_resovled_2011}. The resulting SI is then correlated with the HAADF image, creating a spatial profile of a particular spectral feature, known as LSP mode map. {\bf Figure \ref{SpectraMaps}} shows the schematic operation of a STEM/EELS device in SI mode.  Also included in Figure \ref{SpectraMaps}(b) is a typical EEL spectrum, which consists of three main parts: (i) the zero-loss peak (ZLP) accounts for the large number of electrons that pass through the microscope without interacting with the target and is centered about $\hbar\omega = 0$ eV, (ii) the low-loss region ($\hbar\omega < 50$ eV), which includes LSPs, bulk plasmons, inter- and intra-band transitions, and (iii) the core-loss region ($\hbar\omega > 50$ eV) which probes electronic transitions from inner shells to the conduction band, usually used for el emental mapping and electronic-structure analysis \cite{Pennycook_Atomic_resolution_2009}. The full width at half maximum (FWHM) of the ZLP determines the energy resolution of EELS and bounds its ability to distinguish between features in a spectrum. Thanks to the advances in electron aberration correcting and monochromation of the electron beam \cite{Browning_Monochromators_2006}, modern STEMs can easily reach sub-nanometer spatial resolution with a sub-200 meV energy resolution. To date, STEMs are capable of achieving a spatial resolution of 0.45 nm at 300 keV \cite{Sawada_Resolving_2015} and 0.57 nm  at 200 keV \cite{Dellby_Tuning_high_2014}, while the best energy resolution reported is 9 meV \cite{Krivanek_Vibrational_spectroscopy_2014}. Figure 2(c) shows how an LSP mode map is generated. The most convenient way is to first choose a peak of interest in the spectrum, define a narrow energy window centered at that peak, and then plot the number of counts in the window for each pixel in the SI. Alternatively, instead of plotting the number of counts, an EEL map can also be generated by fitting a Lorentzian to the LSP peak of interest and plotting the area underneath \cite{doi:10.1021/nn5031719}.% Other map generation methods include principle component analysis (PCA) and non-negative matrix factorization (NMF) \cite{Nicoletti_Three_dimensional_2013}.

There are many excellent reviews on the role of STEM/EELS and more broadly on the role of electron microscopy in plasmonics \cite{de2010optical,C3CS60478K}. There the reader will find detailed historical overviews of the theoretical and experimental accomplishments of the field. From a theoretical point of view, LSP and EELS theory can be quite complex \cite{pitarke2007theory}. Rather than repeat what has already been done elsewhere, we narrow our focus and model in detail a prototypical LSP STEM/EELS experiment with hopes of giving researchers new to the field an entry point to the vast body of work that already exists. From the experimental side, we highlight recent and novel applications of EELS that go beyond the simple characterization of LSP modes. These include electron tomography, energy transfer, and quantum-size effects and provide a snapshot of where the field is today. 

\section{Theory of Localized Surface Plasmons and Electron Energy-Loss Spectroscopy}
We begin by building LSP theory and deriving an expression for the EEL probability. The properties of LSPs are greatly influenced by MNP geometry, and it is therefore natural to seek a mathematical description of them in terms of the MNP's surface eigenmodes. While many theoretical approaches exist, classic electrodynamics provides the most straightforward path. The approach outlined in this section can be extended to a variety of geometries; however, to keep the discussion as simple as possible while retaining enough detail to explore the nuances of EELS experiments, we focus solely on spherical MNPs. Moreover, nanospheres are used ubiquitously in nanoscience, and understanding their plasmonic properties is essential for any researcher working in the field.

\subsection{Localized Surface Plasmon Modes of a Metal Nanosphere}

We consider the case of a nanosphere of radius $a$ embedded in a dielectric background with dielectric constant $\varepsilon_b$ and restrict ourselves to particles that are smaller than the diffraction limit of light yet large enough to justify ignoring particle-size quantum effects. This is the quasi-static limit of electrodynamics and amounts to ignoring time delay effects due to the finite speed of light while retaining those due to the inertial response of the MNP's conduction-band electrons. The system's local dielectric function takes the form $\varepsilon({\bf r} ; t-t') = \Theta{(a-r)}\varepsilon(t-t') + \Theta{(r-a)}\varepsilon_b$, where $\Theta$ is the Heaviside step function. For simplicity, we consider $\varepsilon({\bf r} ; t-t') $ to be local in space but not time. In this limit, the response of the system is determined by the time-dependent Poisson equation
\begin{equation}\label{Poisson eqn}
-{\bf\nabla} \cdot \displaystyle \int_{-\infty}^{t} dt'  \varepsilon ( {\bf r}; t - t' ) {\bf\nabla} \Phi ( {\bf r} , t')  = 4 \pi \rho^{\tiny\tiny\textrm{ext}} ( {\bf r}, t),
\end{equation}
where $\Phi( {\bf r} , t) $ is the electric potential due to an external charge density distribution $\rho^{\tiny\textrm{ext}}( {\bf r}, t)$. The electric potential must be finite at the origin and tend to zero as the magnitude of the observation vector $|{\bf r}|\rightarrow \infty$. Furthermore, at the surface of the MNP both the electric potential and the displacement field must be continuous.

To solve Eq.~(\ref{Poisson eqn}), we follow standard procedures \cite{schwinger1998classical} and define the Green function $G({\bf r}, {\bf r}' ; t, t')$ by replacing $\rho^{\tiny\textrm{ext}}$ with a delta function in space and time,
\begin{eqnarray}\label{Green function problem}
-{\bf\nabla}\cdot\displaystyle\int_{-\infty}^{t}dt''\varepsilon({\bf r}; t-t'') {\bf\nabla}  G ( {\bf r}, {\bf r}' ; t'', t' ) = 4\pi \delta( {\bf r} - {\bf r}' )\delta( t - t').
\end{eqnarray}
From a physical point of view this corresponds to finding the potential due to a test charge $q$ located at ${\bf r}'$, which we choose to be external to the MNP. Since we can express any $\rho^{\tiny\textrm{ext}}$ as a sum of point charges, we may invoke the principle of superposition to calculate the total potential by summing Green functions weighted by $\rho^{\tiny\textrm{ext}}$. From a mathematical perspective, this is equivalent to constructing the integral transform that has the dielectric response of the sphere built into its kernel, $G({\bf r}, {\bf r}' ; t, t')$.
 Thus the total potential takes the form
\begin{eqnarray}\label{total potential}
\Phi ({\bf r}, t) =\int dt ' d{\bf r}'G ({\bf r},{\bf r}';t,t')\rho^{\tiny\textrm{ext}} ( {\bf r'}, t' ),
\end{eqnarray}
where the integral is understood to be over all time and space. Intuitively, we expect  $G({\bf r}, {\bf r}' ; t, t')$ to be the sum of two terms, a contribution due to the test charge and a contribution due to the image response of the MNP, and finding its analytic form is the purpose of this section. We begin by taking a Fourier transform of Eq.~(\ref{Green function problem}) in $t-t'$ (assuming $t > t'$), yielding ${\bf\nabla}\cdot\displaystyle\tilde{\varepsilon}\left({\bf r}; \omega \right) {\bf\nabla}  \tilde{G}( {\bf r}, {\bf r}' , \omega ) = -4\pi \delta( {\bf r} - {\bf r}' )$. For the sake of simplicity we choose to treat the metal as a free electron gas by setting $\tilde{\varepsilon}\left(\omega\right) = \varepsilon_{\infty}-\omega_p^2/(\omega^2+i\gamma\omega)$. $\tilde{\varepsilon}\left(\omega\right)$ is parameterized by the bulk plasma frequency $\omega_p$, the electron scattering rate $\gamma$, and $\varepsilon_{\infty}$, the static dielectric response of the ionic background.

Using the completeness relation for spherical harmonics \cite{schwinger1998classical}, $\delta({\bf r} - {\bf r}' )=\sum_{\ell m}r^{-2} \delta{(r-r')} Y_{\ell m}\left(\theta , \phi \right)Y_{\ell m}^* (\theta' , \phi')$, and enforcing the given boundary conditions, we find that the total Green function outside the metal is
\begin{eqnarray}\label{total Green function frequency space}
r > a &:& \tilde{G}({\bf r}, {\bf r}', \omega )=\frac{1}{\varepsilon_b|{\bf r- \bf r'}|} - \sum_{\ell m}\frac{4\pi}{2\ell+1}\frac{\ell \left(\tilde{\varepsilon}\left( \omega \right)/\varepsilon_b -1 \right)}{ \ell \left( \tilde{\varepsilon} \left(\omega \right)+\varepsilon_b\right) +\varepsilon_b}\frac{a^{2\ell+1} }{r'^{\ell+1}r^{\ell+1}}Y_{\ell m}(\theta,\phi)Y_{\ell m}^*(\theta',\phi'),
%r < a &:& \tilde{G}({\bf r},{\bf r}', \omega ) = \sum_{l,m}\frac{(2l+1)\varepsilon_b}{ l ( \tilde{\varepsilon} (\omega)+\varepsilon_b ) +\varepsilon_b} a^{2l+1}\frac{r^{l}}{r'^{l+1}}Y_{\ell m}(\theta,\phi)Y_{\ell m}^*(\theta',\phi'),
\end{eqnarray}
where the spherical harmonics are evaluated at polar and azimuthal observation coordinates $\theta$ and $\phi$, and test charge coordinates $\theta'$ and $\phi'$. As expected it is the sum of two terms, the first corresponding to the test charge, and the second to the MNP's image response. In the limit of $\omega \rightarrow \infty$, we recover the electrostatic Green function for a test charge placed outside of a dielectric sphere with dielectric function $\varepsilon_{\infty}$. Thus we see that polarizing the sphere always lowers the total energy of the system, regardless of the sign of the test charge.

\begin{figure}
\centering
\includegraphics[scale=0.27]{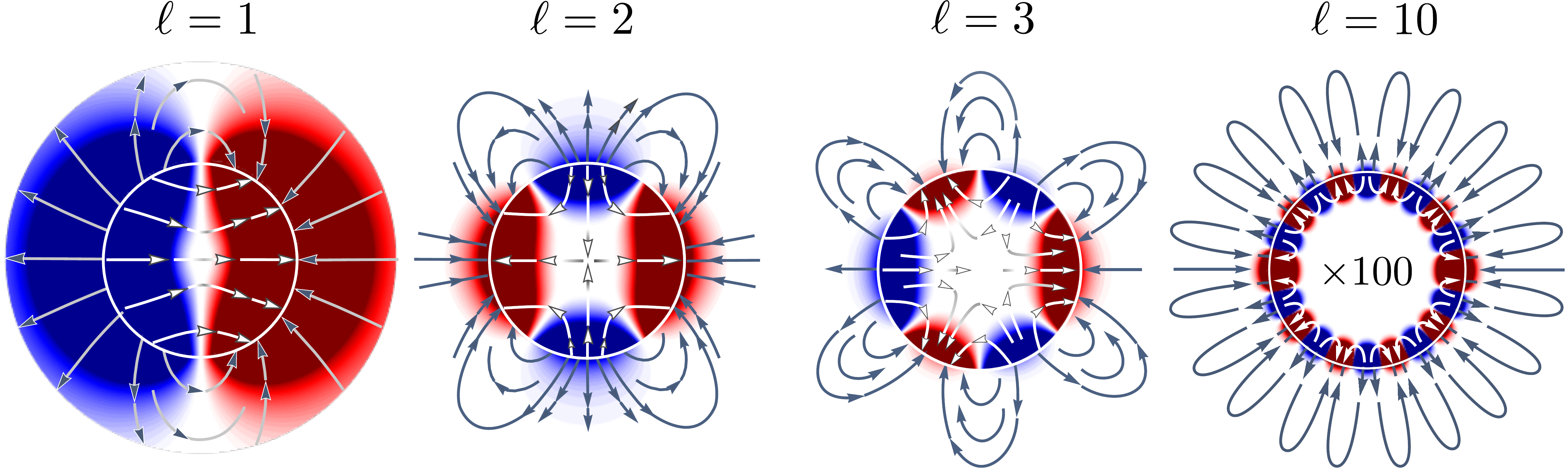}
\caption{Contour plots of LSP electric potential (red and blue signify the sign of the potential) with overlayed electric field lines for $\ell = 1, 2,3,10$ generated by an electron placed on the $x$-axis outside of the sphere. The potential is directly proportional to the surface charge. Bright modes have odd value of $\ell$ and dark modes have even value of $\ell$. $\ell$ gives the number of nodes in the electric field produced by an LSP. Note that the LSP modes are independent of $m$ due to the symmetry of the sphere, giving each $\ell$ mode a $2\ell+1$ degeneracy.}
\label{LSP Modes}
\end{figure}

To see how this connects to the LSP modes of a sphere, we first define the surface response function $\tilde{g}_{\ell m}(\omega)=\ell (\tilde{\varepsilon}( \omega)/\varepsilon_b -1 )/( \ell ( \tilde{\varepsilon} (\omega )+\varepsilon_b) +\varepsilon_b)$, which in the time domain takes the form
\begin{eqnarray}\label{Surface Response Function}
g_{\ell m}(t-t')&-&\frac{\ell(\varepsilon_{\infty}/\varepsilon_b-1)}{\ell (\varepsilon_{\infty}+\varepsilon_b)+\varepsilon_b}\delta(t-t')=4\pi\chi_{\ell m}(t-t')\\  \nonumber
&=& \frac{2 \ell +1}{\ell (\varepsilon_{\infty}+\varepsilon_b)+\varepsilon_b}\frac{\omega_{\ell m}^2}{\sqrt{\omega_{\ell m}^2-(\gamma/2)^2}} e^{-\gamma(t-t')/2} \sin{\left[\sqrt{\omega_{\ell m}^2-(\gamma/2)^2} ~ (t-t') \right]},
\end{eqnarray}
where $\omega_{\ell m}=\omega_p\sqrt{\ell/(\ell(\varepsilon_{\infty}+\varepsilon_b)+\varepsilon_b)}$ are determined by the pole structure of $\tilde{g}_{\ell m}(\omega)$ \cite{pitarke2007theory, ashcroft1976solid}. It is the sum of two terms, an instantaneous response, proportional to a delta function, and an oscillatory term, $\chi_{\ell m}$. The surface response function is one of the primary results of LSP theory because it describes the optical properties of MNPs. This can be seen most explicitly by comparing $g_{\ell m}$ to the response function of a free electron gas,
\begin{eqnarray}\label{Drude time}
\varepsilon(t-t')-\varepsilon_{\infty}\delta (t-t')&=& 4\pi\chi(t-t') \\ \nonumber
&=&\frac{\omega_p^2}{(\gamma/2)}e^{-\gamma (t-t')/2}\sinh\left[\gamma  (t-t')/2\right].
\end{eqnarray}
As described by the Drude model, Eq.~(\ref{Drude time}) indicates that the dynamics of the metal's electrons are overdamped in the bulk, while $g_{\ell m}$ says they oscillate sinusoidally with frequency $\omega_{\ell m}$. This comparison makes it clear that confining the electron gas to a sphere gives the dielectric response of a MNP harmonic oscillator dynamics. Such oscillations are not unique to the existence of a surface. A similar analysis can be carried out for the case of an electron moving through the bulk, yielding a response function $\tilde{g}^{\tiny{B}}(\omega)=(\tilde{\varepsilon}(\omega)-1)/\tilde{\varepsilon}(\omega)$, which is singular at $\omega_p/\sqrt{\varepsilon_\infty}$, the well known bulk plasmon resonance.

By analogy to Eq.~(\ref{Drude time}), we see that $\chi_{\ell m}$ is the multipolar MNP susceptibility which can be used in conjunction with Eq.~(\ref{total potential}) to define the dynamic part of the induced potential,
\begin{eqnarray}
%\nonumber
\Phi^{\tiny\textrm{ind}}({\bf r},t)
% &=& V\sum_{\ell m} \frac{3}{2\ell + 1} \frac{a^{\ell-1}}{r^{\ell + 1}}Y_{\ell m}(\theta,\phi)\int dt'd{\bf r'} \chi_{\ell m}(t-t')\rho^{\tiny\textrm{ext}}({\bf r'},t')\frac{a^{\ell-1}}{r'^{\ell+1}}Y^*_{\ell m}(\theta',\phi') \\ 
\label{Induced Phi SH}
= -V\sum_{\ell m} f_{\ell m}({\bf r})\int dt'd{\bf r'} \chi_{\ell m}(t-t')\rho^{\tiny\textrm{ext}}({\bf r'},t')f_{\ell m}^{*}({\bf r'}).
\end{eqnarray}
Here $V=4\pi a^3/3$ is the particle volume and $f_{\ell m}({\bf r})=\sqrt{12\pi/(2\ell +1)}a^{\ell-1}r^{-\ell-1}Y_{\ell m}(\theta,\phi)$. Given the form of Eq.~(\ref{Induced Phi SH}), it is convenient to define the induced multipole moments
\begin{eqnarray}\label{MM}
q_{\ell m}(t)=V\int dt' \chi_{\ell m}(t-t')\int d{\bf r'}\rho^{\tiny\textrm{ext}}({\bf r'},t')f_{\ell m}^{*} ({\bf r'}).
\end{eqnarray}
%The instantaneous electrostatic response of the system  has been ignored since it only serves to lower the total energy by an overall constant. 
Thus from analysis of the Green function, we are now able to understand what is meant by localized surface plasmons. They are the collective response of the metal's free carriers, whose nature has been fundamentally altered by the presence of a confining surface. The LSPs of the sphere have resonance frequency $\omega_{\ell m}$ independent of $\rho^{\tiny\textrm{ext}}$, meaning that LSPs are intrinsic to the MNP. The angular momentum number $\ell$ orders the modes by increasing energy and number of nodes in the surface charge distribution. LSPs with odd $\ell$ values are bright, while those with even $\ell$ values are dark. The information encoded in the value of $\ell$ is also evident in the spatial profile of the LSP's electric field. {\bf Figure (\ref{LSP Modes})} shows the potentials and electric fields for the $\ell = 1,2,3$, and $10$ modes.

We see from Eq.~(\ref{MM}) that only the amplitude is determined by $\rho^{\tiny\textrm{ext}}$, meaning that the subset of LSPs accessible is entirely determined by its form. For example, if we choose $\rho^{\tiny\textrm{ext}}=q\delta({\bf r}-{\bf r}')$ and place the source charge $q$ far from the MNP, we find that only the $\ell=1 $ term contributes to Eq.~(\ref{MM}). When combined with the instantaneous response, we find that to first order a nanosphere in free space ($\varepsilon_b=1$) is an electric dipole of energy $\hbar\omega_{1m}=\hbar\omega_p/\sqrt{\varepsilon_{\infty}+2}$ and a frequency-space dipole moment
\begin{eqnarray}\label{Dipole}
\tilde{p}(\omega)=\sum_m\tilde{q}_{1m}(\omega)=a^3\frac{\tilde{\varepsilon}(\omega)-1}{\tilde{\varepsilon}(\omega)+2}\left[\frac{\partial}{\partial r}\frac{q}{|{\bf r}-{\bf r'}|}\right]_{{\bf r}= {\bf 0}}=\tilde{\alpha}_{\tiny\textrm{CM}}(\omega)|{\bf E}_{\tiny\textrm{local}}|.
\end{eqnarray}
Here, $\tilde{\alpha}_{\tiny\textrm{CM}}(\omega)$ is implicitly defined as the Clausius$-$Mossotti  polarizabilty \cite{frohlich1949theory} of a sphere and $|{\bf E}_{\tiny\textrm{local}}|$ is the electric field magnitude of the source charge evaluated at the origin. 
\subsection{Electron Energy-Loss Spectroscopy Theory}
We now show that the EEL probability measured in STEM/EELS experiments is directly related to the Green function calculated above. Intuitively, this should be expected since $G({\bf r}, {\bf r}'; t-t' )$ is calculated by placing a stationary charge in the vicinity of the MNP, while in STEM/EELS the MNP interacts with an electron on the trajectory, ${\bf x}(t)$. This motivates us to choose $ \rho^{\tiny\textrm{ext}}=q\delta({\bf r}-{\bf x}(t))$, and derive the electron's equation of motion in a background potential determined by Eq.~(\ref{Induced Phi SH}). More specifically, we want the back-force due to the image response of the MNP at the present location of the electron. Setting the observation coordinate ${\bf r}  = {\bf x}(t)$ and taking the integral over ${\bf r}'$ we obtain the equation of motion,
\begin{eqnarray}\label{electron eom}
m_e\ddot{{\bf x}}(t)=e^2\sum_{\ell m}\left[{\bf \nabla}f_{\ell m}({\bf r})\right]_{{\bf r}= {\bf x}(t)}V\int_{-\infty}^{t} dt ' \chi_{\ell m}(t-t')f_{\ell m}^{*}({\bf x}(t')),
\end{eqnarray}
for $q = -|e|$ and electron mass $m_e$. The analysis of the above nonlinear integro-differential equation can be simplified by exploiting the harmonic oscillator dynamics of the multipole moments defined in Eq.~(\ref{MM}). 
%This allows us to define an effective LSP displacement coordinate,
% $Q_{\ell m}(t)=V\int_{-\infty}^{t} dt ' \chi_{\ell m}(t-t')f_{\ell m}({\bf x}(t')$, that is being driven by a force $F_{\ell m}(t) =e^2f_{\ell m}({\bf x}(t))$. and has an associated Green function $V\chi_{\ell m}(t-t')$. 
This allows us to recast Eq.~(\ref{electron eom}) as a set of coupled equations of motion where an effective LSP coordinate, $Q_{\ell m} = V \int dt' \chi_{\ell m}(t-t')f_{\ell m}^{*}({\bf x}(t'))$, is driven by a force, $F_{\ell m}(t) =e^2f_{\ell m}({\bf x}(t))$. We find 
\begin{eqnarray}\label{system x}
 &\ddot{{\bf x}}(t)=e^2\sum_{\ell m}\left[{\bf \nabla}f_{\ell m}({\bf r})\right]_{{\bf r}= {\bf x}(t)}Q_{\ell m}/m_e \\\label{system Q}
 &\ddot{Q}_{\ell m} + \gamma\dot{Q}_{\ell m}+\omega_{\ell m}^2 Q_{\ell m}=F_{\ell m}(t)/m_{\ell m}, 
\end{eqnarray}
with $m_{\ell m}=4\pi e^2(2\ell+1)/\left[\omega_{\ell m}^2V(\ell(\varepsilon_{\infty}+\varepsilon_b)+\varepsilon_{b})\right]$, the LSP effective mass. In the limit of $\gamma \to 0$, the total energy is
\begin{eqnarray}\label{Hamiltonian}
H = \frac{{\bm \wp}^2}{2m_e}+\sum_{\ell m}\left[\frac{P_{\ell m}^2}{2m_{\ell m}}+\frac{1}{2}m_{\ell m}\omega_{\ell m}^2 Q_{\ell m}^2\right]-\sum_{\ell m}Q_{\ell m}F_{\ell m}({\bf x}),
\end{eqnarray}
which is the system's Hamiltonian, from which we can compactly derive Eqs.~(\ref{system x}) and~(\ref{system Q}). Here, ${\bm \wp}$ is the electron's momentum conjugate to ${\bf x}$ while $P_{\ell m}$ is conjugate to $Q_{\ell m}$. Eq.~(\ref{Hamiltonian}) defines a mechanical system that has been parameterized to mimic the dynamics of the electron-MNP system. 

Using Eq.~(\ref{Hamiltonian}), we are free to calculate corrections to the electron's trajectory due to the nonlinear coupling to the LSPs. Such dynamics are important when the energy of the electron is commensurate with $\hbar \omega_{\ell m}$. Fortunately, STEM electrons have energy on the keV scale, three orders of magnitude greater than that of LSPs. In this limit, the electron beam can be viewed as a bath degree of freedom, effectively providing the system an infinite source of energy, and we can safely assume that the electron moves at constant velocity, ${\bf v}=v\hat{{\bf k}}$. This is the well-known no-recoil limit and amounts to ignoring ${\bm \wp}^2/2m_e$ in Eq.~(\ref{Hamiltonian}) and setting ${\bf x}(t) = x_0\hat{{\bf i}}+y_0\hat{{\bf j}}+(\wp_z /m_e)t\hat{{\bf k}}$, where $x_0$ and $ y_0$ set the beam position. This yields
\begin{eqnarray}
%\label{Classical Hamiltonian}
%H&=&\sum_{\ell m}\left[\frac{P_{\ell m}^2}{2m_{\ell m}}+\frac{1}{2}m_{\ell m}\omega_{\ell m}^2 Q_{\ell m}^2\right]-\sum_{\ell m}Q_{\ell m}F_{\ell m}({\bf x}) \\
\label{QM Hamiltonian}
H=\sum_{\ell m}\hbar\omega_{\ell m}\left(b_{\ell m}^{\dag}b_{\ell m}+\frac{1}{2}\right)-\sum_{\ell m}\sqrt{\frac{\hbar}{2m_{\ell m}\omega_{\ell m}}}F_{\ell m}\left(\frac{\wp_z}{m_e}t;R_0\right)\left(b_{\ell m}+b_{\ell m}^{\dag}\right),
\end{eqnarray}
where the impact parameter $R_0 = \sqrt{x_0^2+y_0^2}$. We have quantized the system by defining bosonic raising and lowering operators $b_{\ell m}$ and $b_{\ell m}^{\dag}$, leaving an interaction Hamiltonian $H_{\tiny\textrm{int}}$ that couples the LSP modes to the electron beam.

In this description of the MNP-STEM electron system, an initial LSP state prepared in the distant past, $|\psi(t=-\infty)\rangle$, time evolves in the interaction picture with time evolution operator $\mathcal{U}(t,-\infty)=\exp{\left\{(-i/\hbar)\int_{-\infty}^{t}dt' H_{\tiny\textrm{int}}(t')\right\}}$, giving
\begin{eqnarray}
\label{Time evolution}
|\psi(t)\rangle
% &=& \exp\left\{\frac{i}{\hbar}\int_{-\infty}^{t}dt' \sum_{\ell m}\sqrt{\frac{\hbar}{2m_{\ell m}\omega_{\ell m}}}F_{\ell m}(t';v,b)(a_{\ell m}(t') + a_{\ell m}^\dagger(t'))\right\}|\psi(-\infty)\rangle \\
= \prod_{\ell m}\exp\left\{\beta_{\ell m}(t)b^\dagger_{\ell m} - \beta_{\ell m}^*(t)b_{\ell m}\right\}|\psi(-\infty)\rangle. \label{Coherent State}
\end{eqnarray}
Here, $\beta_{\ell m}(t) = i\int_{-\infty}^{t}dt' \sum_{\ell m}F_{\ell m}((\wp/m_e)t;R_0)e^{i\omega_{\ell m}t'}/\sqrt{{2\hbar m_{\ell m}\omega_{\ell m}}}$, and we have used $b_{\ell m}(t)=b_{\ell m}e^{-i\omega_{\ell m}t}$ \cite{devreese2013elementary}. If we assume that the LSP modes are initially in the ground state, Eq.~(\ref{Coherent State}) is a definition of a multimode coherent state with time dependent amplitudes $\beta_{\ell m}(t)$. Further recognizing that the probability for the electron to lose energy to the LSP modes is equal to the probability for the LSP modes to be found in an excited state gives us a route to calculate the EEL spectrum \cite{devreese2013elementary, PhysRevLett.55.1526}. In the steady-state limit ($t\rightarrow\infty$), after the electron and MNP have stopped interacting, the probability of finding the MNP with a particular set of LSP mode occupation numbers is,
\begin{eqnarray}\label{Poisson}
P(n_{10}, n_{11}, ..., n_{\ell m},...)=|\langle n_{10}, n_{11}, ..., n_{\ell m},...|\psi(\infty)\rangle|^2= \prod_{\ell m} \frac{|\beta_{\ell m}(\infty)|^{2n_{\ell m}}}{n_{\ell m}!}e^{-|\beta_{\ell m}(\infty)|^2},
 \end{eqnarray}
that is, the occupation numbers obey Poisson statistics with average $|\beta_{\ell m}(\infty)|^2$. Interestingly, this implies that the electron beam has nonzero probability of exciting multiple plasmons in a single mode, a result that agrees with observation of multiple loss peaks \cite{barwick2009photon}. For simplicity, however, we restrict ourselves to single quanta loss events, and write the total probability of exciting a single plasmon in mode $\ell, m$ as
\begin{equation}\label{Prob1}
\begin{split}
P_{\ell m}(n_{\ell m}=1) &= |\beta_{\ell m}(\infty)|^2 e^{-|\alpha_{\ell m}(\infty)|^2} \approx  |\beta_{\ell m}(\infty)|^2 \\
&= \frac{2e^2\omega_{\ell m}a}{\hbar v^2}\left(\frac{a\omega_{\ell m}}{v}\right)^{2\ell} \frac{2\ell +1}{(\ell + m)!(\ell -m)!}\frac{K_m^2\left(\left|\frac{\omega_{\ell m} R_0}{v}\right|\right)}{\ell(\varepsilon_\infty + \varepsilon_b)+\varepsilon_b},
\end{split}
\end{equation}
where we assume $|\beta_{lm}(\infty)|^2 \ll 1$ since most STEM electrons pass the MNP without interacting, and we have calculated the Fourier transform of $F_{\ell m}(t)$, and $K_m$ is the modified Bessel function of the second kind \cite{ferrell1987analytical}.
\begin{figure}
\centering
\includegraphics[scale=0.3]{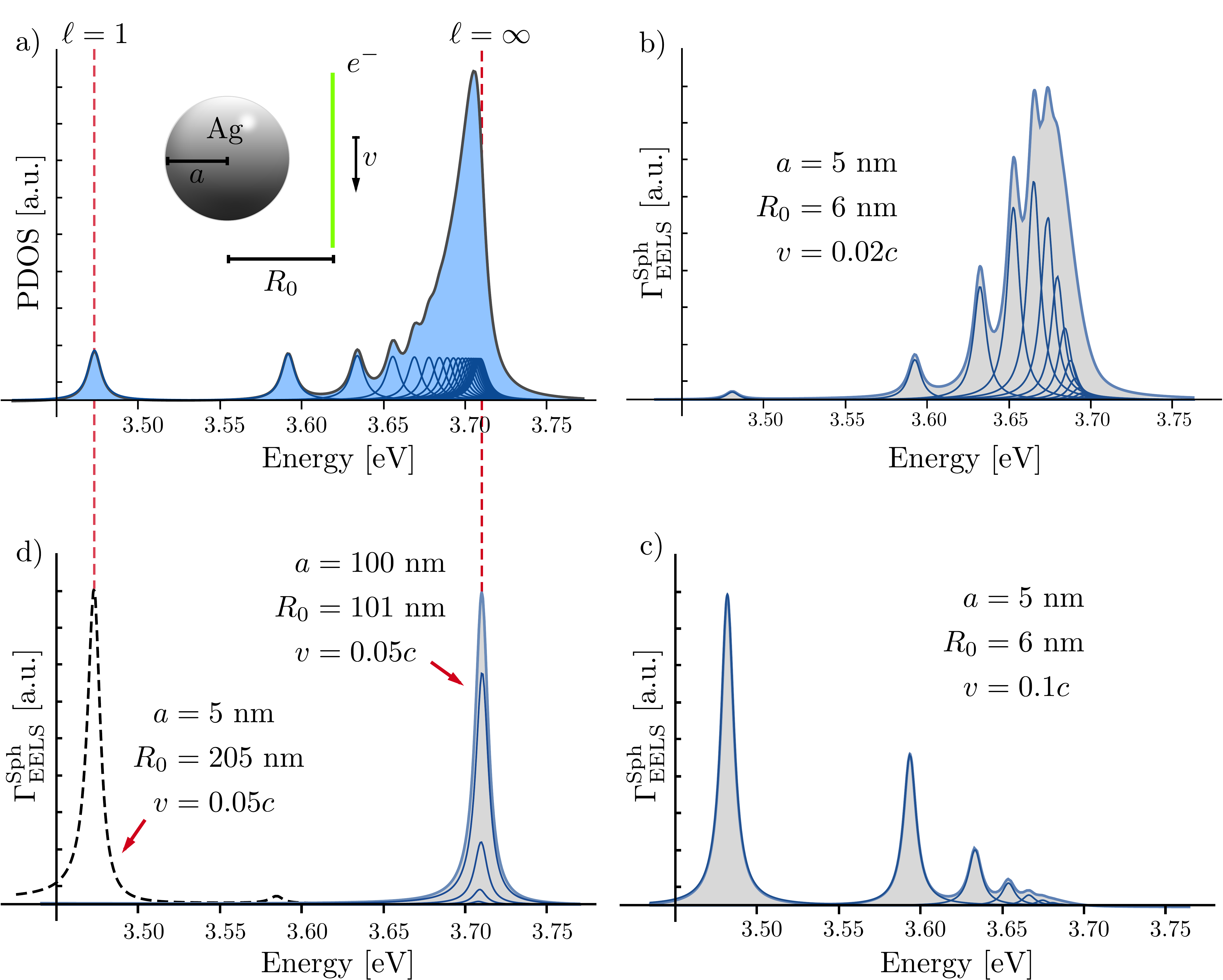}
\caption{Plasmonic density of states (PDOS) and EEL probability per unit energy of Ag nanospheres. All calculations assume a Drude model with $\varepsilon_{\infty}=5.7$, $\omega_p=9.3$ eV and $\gamma = 0.01$ eV, and all plots were made by summing the appropriate expression for the first $\ell=20$ terms. (a) The Ag nanosphere PDOS exhibits bunching near $3.72$ eV, the SPP energy of a Ag/vacuum plane interface, which is also the $\ell\to\infty$ limit of $\omega_{\ell m}$. (b)-(d) EEL probability density for Ag nanosphere excited by a passing electron for varying size, speed (measured relative to the speed of light, $c$), and impact parameter. Depending on the relative values of these parameters, the electron beam will preferentially interact with a subset of modes found in the PDOS. Note the dominance of the dipole $\ell=1$ mode when $R_0=205$ nm and $a=5$ nm, in contrast to the dominance of the SPP type mode when $a= 100$ nm and $R_0-a=1$ nm. }
\label{Hybridization}
\end{figure}

Eq.~(\ref{Prob1}) can be used to construct the probability density associated with exciting a single plasmon in any of the MNP modes by weighting the $\ell m$th element of the plasmonic density of states (PDOS) per unit frequency $\rho(\omega) =\sum_{\ell m}\rho_{\ell m}(\omega) =  \sum_{\ell m}\delta(\omega-\omega_{\ell m})$ by $P_{\ell m}(n_{\ell m}=1)$, yielding

\begin{equation}\label{EEL Probability}
\begin{split}
\Gamma^{\tiny{\textrm{Sph}}}_{\tiny\textrm{EELS}}(\omega) &= \sum_{\ell m} P_{\ell m}(n_{\ell m} = 1)\rho_{\ell m}(\omega) \\
&= \sum_{\ell m} \frac{3e^2}{\hbar \pi^2 a^2 v^2}\left(\frac{a\omega}{v}\right)^{2\ell}\frac{K_m^2\left(\left|\frac{\omega R_0}{v}\right|\right)}{(\ell + m)!(\ell -m)!}\left[\frac{V\pi(2\ell + 1)\omega_{\ell m}^2}{\ell(\varepsilon_\infty + \varepsilon_b)+\varepsilon_b}\delta(\omega^2 - \omega^2_{\ell m})\right]\\
&\stackrel{\gamma\neq0}{\longrightarrow}  \sum_{\ell m} \frac{3e^2}{\hbar \pi^2 a^2 v^2}\left(\frac{a\omega}{v}\right)^{2\ell}\frac{K_m^2\left(\left|\frac{\omega R_0}{v}\right|\right)}{(\ell + m)!(\ell -m)!} \textrm{Im}\left\{V\tilde{\chi}_{\ell m}(\omega)\right\}
\end{split}
\end{equation}
where we have used the relations $\delta(\omega - \omega_i) + \delta(\omega+\omega_i) = 2\omega_i\delta(\omega^2 - \omega_i^2)$ and $\textrm{Im}(\omega-\omega_i\pm i\gamma)^{-1}=\mp\pi\delta(\omega-\omega_i)$. Note, that since $\textrm{Im}\{V\tilde{\chi}\}=4\pi\textrm{Im}\{\tilde{\alpha}\}/3$, where $\tilde{\alpha}$ is the generalized polarizabilty of the MNP, we recover the expression found in reference \cite{de2010optical}.

In the nonrelativistic limit, $v < 0.1c$, the EEL probability $P_{\ell m}$ is the integral of $\Gamma_{\tiny\textrm{EELS}}(\omega)$ over frequency bins of width determined by the spectral resolution of the STEM. Eq.~(\ref{EEL Probability}) further shows that EELS probes the PDOS with weights dependent on the velocity and location of the electron beam, a result we may have expected based on intuition developed with the Green function. Somewhat surprisingly, the EEL probability also depends on the radius of the nanosphere even though, in the quasistatic limit, the PDOS does not. Eq.~(\ref{EEL Probability}) has a complex dependence on these three experimentally tunable parameters, and different choices of beam position, velocity, and nanosphere size often yield EEL spectra dominated by different LSP modes \cite{PhysRevLett.55.1526}. Examples of this are shown explicitly in {\bf Figure 4} where the loss probabilities of Ag nanospheres are plotted alongside the PDOS. Although the peaks do not move from spectrum to spectrum, their heights can change dramatically. This high dimensional parameter space is a unique feature of EELS, and for this reason, great care must be taken when interpreting EEL spectra. 

\subsection{Numerical Methods}
Although nanospheres offer a simple starting point for understanding EELS, modeling experiment often requires a quantitative description of more complicated nanostructures. As a result, numerical techniques for solving Maxwell's equations sourced by the electron beam have become an indispensable tool for studying EELS. Broadly speaking, these methods approximate the nanostructure's continuous degrees of freedom by a discrete set, thereby converting Maxwell's equations into a set of algebraic relations that can be solved numerically. This discretization procedure is not unique, and a number of approaches exist. An in-depth comparison of different methods is beyond the scope of this review, and we instead focus introducing reference materials for three of the most popular methods in EELS.

The discrete dipole approximation (DDA) is due to Purcell and Pennypacker \cite{purcell1973scattering}, who showed that the scattering properties of a target can be approximated by treating it as a finite set of coupled dipoles, which are polarized by external sources. For EELS purposes, the electron-driven discrete dipole approximation ($e$-DDA) \cite{bigelow2012characterization} is an open source software package that incorporates the electron beam into the DDA based scattering code DDSCAT \cite{draine2013user}. This approach has a track record of excellent agreement with experiments \cite{geuquet2010eels,yang1995discrete,bigelow2013signatures,li2015spatially}, and its efficiency and implementation is a topic of current research \cite{draine1993beyond,draine1994discrete,guillaume2013efficient}.

Another approach using the boundary element method (BEM) \cite{hall1994boundary} is implemented in the Matlab based package metal-nanoparticle BEM (MNPBEM) \cite{2012mnpbem,hohenester2014simulating}. Here, Maxwell's equations are written in terms of surface integrals, and the effect of the electron beam is computed by considering the response of a discretized surface charge distribution \cite{de1997numerical,de1998relativistic,de2002retarded}. This approach has had wide success as well \cite{MAM:9289869,horl2013tomography}, and MNPBEM is rapidly being improved to include other EELS related experiments \cite{waxenegger2015plasmonics}.

The finite-difference time-domain method (FDTD) method is based on the Yee algorithm \cite{yee1966numerical}, which uses staggered space and time grids for electric and magnetic fields respectively, ensuring that this discretization satisfies Gauss' law \cite{taflove2005computational}. FDTD is the most widely used numerical approach to Maxwell's equations, and it is well documented  \cite{taflove2005computational}. There are many open source and commercial implementations of FDTD and some have started to model EELS explicitly \cite{oskooi2010meep,cao2015electron}. 

Researchers have also modeled EELS using T-matrix approaches \cite{matyssek2012t}, generalized multipole techniques \cite{kiewidt2013numerical}, discontinuous Galerkin methods \cite{matyssek2011computing}, and density functional theory (DFT) \cite{dudarev1998electron}, all with various degrees of success. It is important to note that, for a particular problem, some methods are better than others, and when deciding which to use, it is often illuminating to compare results from several methods. 

\section{Localized Surface Plasmon Mode Mapping}
Traditional EELS acquires a spectrum at a fixed beam position leaving loss energy as the only independent variable. While this approach has been used successfully for decades, it is not able to visualize the spatial profile of the electric field of plasmonically active MNPs. Alternatively, as was illustrated in Figure \ref{SpectraMaps}, the SI mode links the loss spectrum to the electron beam's spatial coordinates. The data can be re-interpreted to visualize the 2-D spatial profile of the loss probability at a given energy. Figure \ref{SpectraMaps}(c) illustrates the generation of a dipolar LSP mode map from a spectrum image.

\begin{figure}
\centering
\includegraphics[scale=0.55]{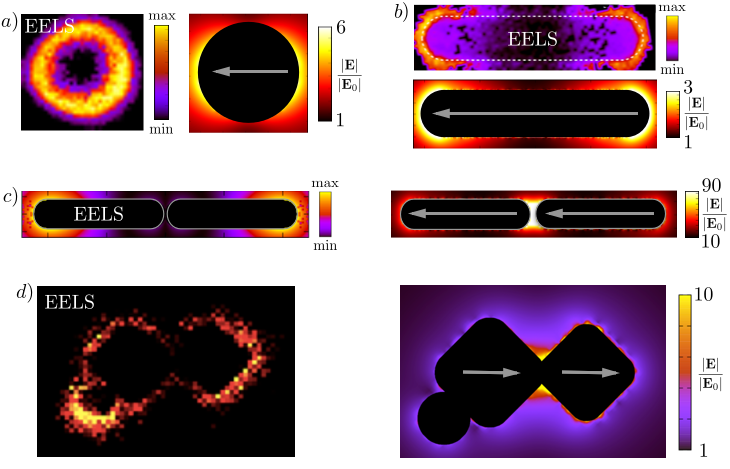}
\caption{LSP mode map compared to the plane-wave induced electric field magnitude at the same loss-energy value for: (a) the dipole plasmon of a Ag truncated nanosphere. Note the radial symmetry of the mode map in contrast to the nodal structure seen in the electric field plot. Reproduced with permission from Reference \cite{Li_Examining_substrate_2015}. Copyright 2015 American Chemical Society. (b) the dipole plasmon of a Ag nanorod. The mode map and electric field magnitude share a similar spatial profile due to the reduced symmetry of the system. (c) the collective dipole plasmon of a Ag nanorod dimer. Note the similarity between the two ends of the rod in contrast to anti-correlation of the two in the junction. Reproduced with permission from Reference \cite{bigelow2012characterization}. Copyright 2012 American Chemical Society. (d) the collective dipole plasmon of a silver nanoparticle aggregate that serves as a plasmonic antenna for single-molecule SERS. While the locations of high electric field strength and high EEL probability are not spatially co-located, Mirsaleh-Kohan et al. demonstrated that they are correlated nonetheless \cite{Mirsaleh_Single_molecule_2012} in agreement with prediction \cite{bigelow2012characterization}.}
\label{Mode Maps pitfalls}
\end{figure}

As noted by several authors \cite{PhysRevLett.103.106801,de2008probing}, some caution must be taken when interpreting mode maps due to the nature of the EELS experiment. In EELS, the electron beam is the excitation source as well as the measurement probe, and therefore EELS is only proportional to the \textit{magnitude} of the electron-induced electric field \textit{at the location} of that same electron. Effectively, this means that mode maps do not directly correlate to the LSP's electric field, but rather to a spatial average of its electric field magnitude in the plane normal to the electron beam's trajectory. This is true at all points in space except where excitation of the mode is disallowed due to electron-beam imposed selection rules, where the loss probability is zero. For certain geometries, such as particles with edges and corners, there is no qualitative difference between mode maps and the electric field. But in systems with a high degree of symmetry, such as spheres, or multi-particle systems with junctions, some care must be taken since not all regions in space with zero loss probability correlate with zero electric field. {\bf Figure \ref{Mode Maps pitfalls}} compares loss probability maps to electric field magnitude plots of selected nanoparticle systems.

Though the basic science needed to create mode maps has been around for decades, it was not until 2007 that Nelayah et al. were able to image the LSP modes of Ag nanoprisms \cite{Nelayah_Mapping_2007}. This technique was quickly improved upon and applied to a variety of particles such as nanorods \cite{Guiton_Correlated_optical_2011, Rossouw_Multipolar_plasmonic_2011, Alber_Visualization_2011}, nanodecahedra \cite{Myroshnychenko_Plasmon_spectroscopy_2012}, nanodisks \cite{Schmidt_Morphing_2014, Schmidt_breathing_modes_2012}, nanocubes \cite{Mazzucco_Ultralocal_modification_2012, Iberi_Resonance_Rayleigh_2014}, truncated nanospheres \cite{Li_Examining_substrate_2015} and multi-nanoparticle systems \cite{Koh_lithographically_defined_2011, Barrow_nanoparticle_chains_2014, Mirsaleh_Single_molecule_2012, Alber_nanowire_dimers_2012}. Mode maps allow researchers to experimentally visualize the near-field character of LSPs and correlate the near-field profile of LSPs to nanoparticle geometry, something that previously required the use of theory. Despite these successes, due to the limitations of 2-D mode maps, the behavior of the system in the direction parallel to the electron beam remains unknown in most EELS experiments. However, in some situations, such as that of a nanoparticle sitting on a dielectric substrate, two dimensions is insufficient. Specifically, substrates are known to induce LSP mode splitting, a phenomenon often accompanied by plasmon localization in the direction normal to the substrate, which can be visulized only with the development of 3-D mode mapping technology. Given that all systems studied in the STEM must be placed on a substrate, this effect can play a central role in the analysis of STEM/EELS experimental data. 

Substrate-induced mode splitting can be interpreted  as the mixing of initially uncoupled LSP modes via the dielectric response of the substrate. This mechanism was used to explain the EELS spectra of a nanocube on a dielectric substrate, where the dipole and quadrupole corner modes renormalize into vacuum and substrate localized dipole modes \cite{Sherry_silver_nanocubes_2005, Zhang_Fano_resonances_2011, Ringe_nanocubes_2010}. Due to the limitations of 2-D LSP mode mapping, this effect could not be directly visualized and therefore could only be inferred through the aid of theory and simulation.

\begin{figure}
\includegraphics[scale=0.5]{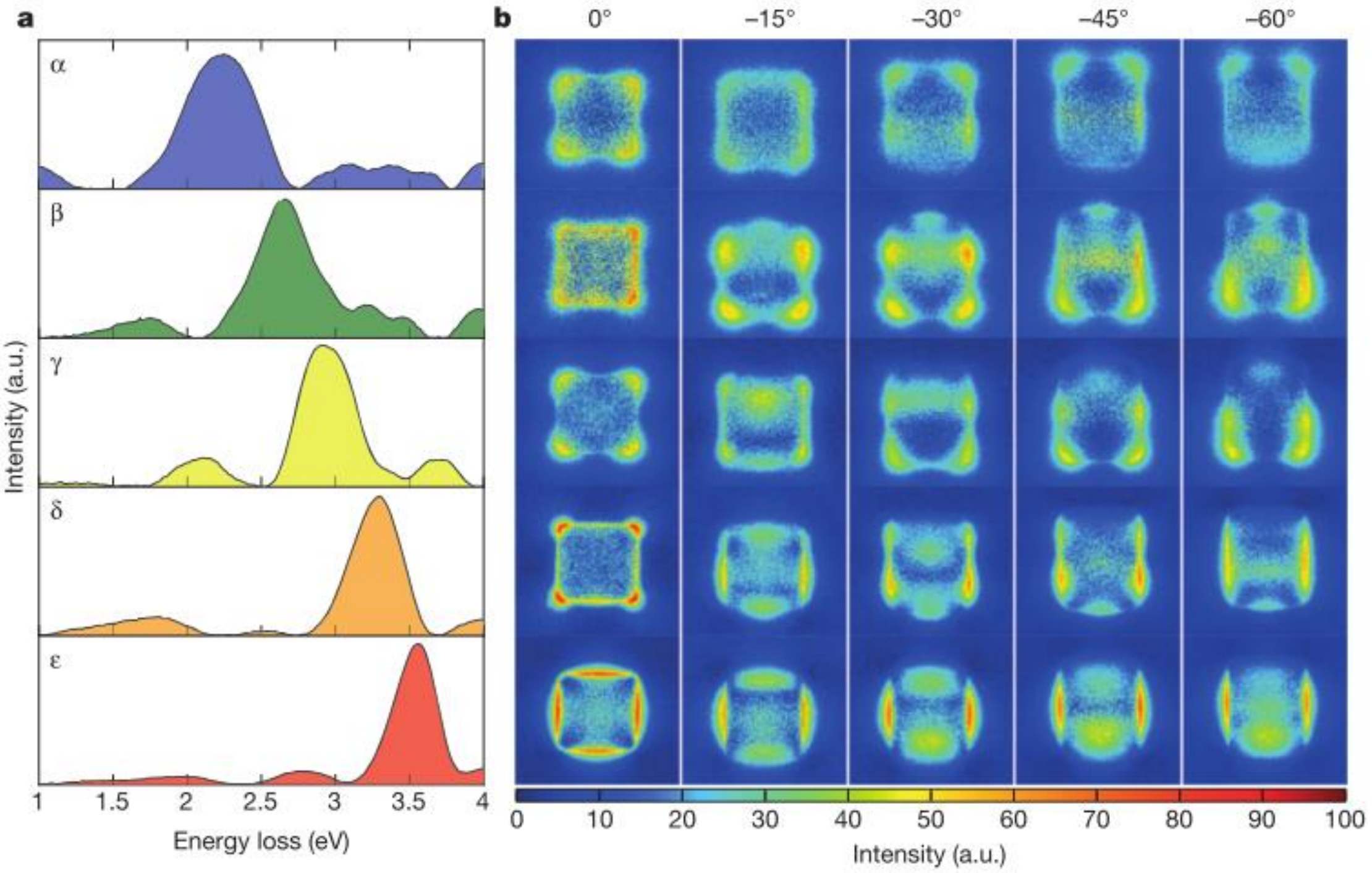}
\caption{LSP components and tomographic EELS maps of a Ag nanocube. (a) Spectral components for five major surface excitations ($\alpha$, $\beta$, $\gamma$, $\delta$ and $\varepsilon$) produced by the NMF method, in order of increasing energy loss. (b) Normalized EELS maps corresponding to the five NMF components shown on the left. Reproduced with permission from Reference \cite{Nicoletti_Three_dimensional_2013} . Copyright 2013 Nature Publishing.}
\label{Midgley}
\end{figure}

Recently, Midgley and co-workers \cite{Nicoletti_Three_dimensional_2013} showed how 2-D LSP mode mapping can be improved upon by combining EELS with electron tomography. They showed that by collecting EELS spectra on a tilted platform over many angles, it was possible to construct 3D mode maps. To showcase the usefulness of this technique, they studied a Ag nanocube on a silicon nitride substrate and acquired EELS spectral images for every 15$^{\circ}$ of tilt. This allowed them to build a 3-D spatial profile of the nanoparticle's LSP modes as well as directly image substrate-induced mode splitting in Ag nanocubes. They found that the LSP modes of the nanocube are, as predicted by theory, quite complex due to the degeneracy of modes with differing spatial profiles. The energetic evolution of nanocube LSP modes does not follow the usual low energy dipole to higher energy multipole trend found in simpler geometries. The modes of a nanocube can be classified into sets of corner, edge and face modes, with each set having its own energy ordering that obeys the usual multipolar progression. The effect of the substrate is to create new sets of modes that are linear combinations of the free-space modes. Midgley et. al. defines five sets of modes, grouped according to their energetic overlap as $\alpha$, $\beta$, $\gamma$, $\delta$, and $\varepsilon$. Mode maps for each set are shown from five viewing angles in {\bf Figure \ref{Midgley}}, highlighting the importance of tomographic imaging in the characterization of complex geometries.

\section{Energy Transfer}
One of the open challenges in plasmonics is to develop a complete understanding of competing energy transfer mechanisms between a MNP and its environment. In particular, the use of plasmonic particles in conventional solar energy harvesting devices, which are based on the use of semiconductors with optical frequency band-gaps, has generated a significant amount of interest in recent years \cite{catchpole2008plasmonic}. While traditional solar devices are becoming increasingly more efficient and affordable, the underlying semiconductor physics of these devices limits their effectiveness to a relatively narrow band of the available solar spectrum. Searching to overcome these limitations, researchers have used the unique optical properties of LSPs to improve energy-harvesting efficiency by embedding MNPs into solar devices. 

The overall efficiency of the device could be enhanced via one or more of the following mechanisms: (1) Photon-LSP scattering, which increases the likelihood of photon absorption by the semiconductor \cite{Schaadt_Enhanced_semiconductor_2005, Nakayama_solar_cells_2008}; (2) Plasmon-induced resonance energy transfer (PIRET), which occurs between an LSP and the interband transition dipole moment of the semiconductor \cite{Li_plasmonic_photocatalysts_2012, Cushing_energy_harvesting_2013}; (3) Hot electron injection generated through the decay of an LSP into an electron-hole pair, a process known as direct electron transfer (DET) \cite{Li_plasmonic_photocatalysts_2012, Cushing_Photocatalytic_activity_2012, Claverp_hot_electron_2014, Brongersma_hot_carrier_2015, Hoggard_electron_transfer_2013}. PIRET requires spectral overlap between the LSP emission and the semiconductor absorption \cite{Li_plasmonic_photocatalysts_2012, Cushing_Photocatalytic_activity_2012, Cushing_energy_harvesting_2013}, while DET is only available when the hot electron is energetic enough to overcome the Schottky barrier at the interface \cite{Li_plasmonic_photocatalysts_2012, Li_water_splitting_2013, Cushing_Photocatalytic_activity_2012, Cushing_energy_harvesting_2013, Claverp_hot_electron_2014, Brongersma_hot_carrier_2015, Hoggard_electron_transfer_2013}. Mechanism (1) is only effective at energies above the band-gap, whereas it has been suggested that mechanism (2) and (3) allow for energy transfer to occur below and above the band-gap energies \cite{Li_plasmonic_photocatalysts_2012, Li_water_splitting_2013, Cushing_Photocatalytic_activity_2012, Cushing_energy_harvesting_2013, Seh_hydrogen_generation_2012}, thereby widening the amount of the solar spectrum accessible to the device \cite{Seh_hydrogen_generation_2012, Tian_photoelectrochemistry_2004, Furube_gold_nanodots_2007, Liu_water_splitting_2011, Desario_water_splitting_2013, Linic_efficient_conversion_2011, DuChene_solar_photocatalysis_2014, Kawawaki_photocurrents_2013}. While the unique optical properties of plasmonic nanoparticles have been harnessed to greatly enhance the sensitivity of Raman and fluorescence spectroscopies \cite{jeanmaire1977surface}, achieving comparable gains in plasmon-enhanced photovoltaics and photocatalysis remain elusive. 
\begin{figure}
\centering
\includegraphics[scale=0.5]{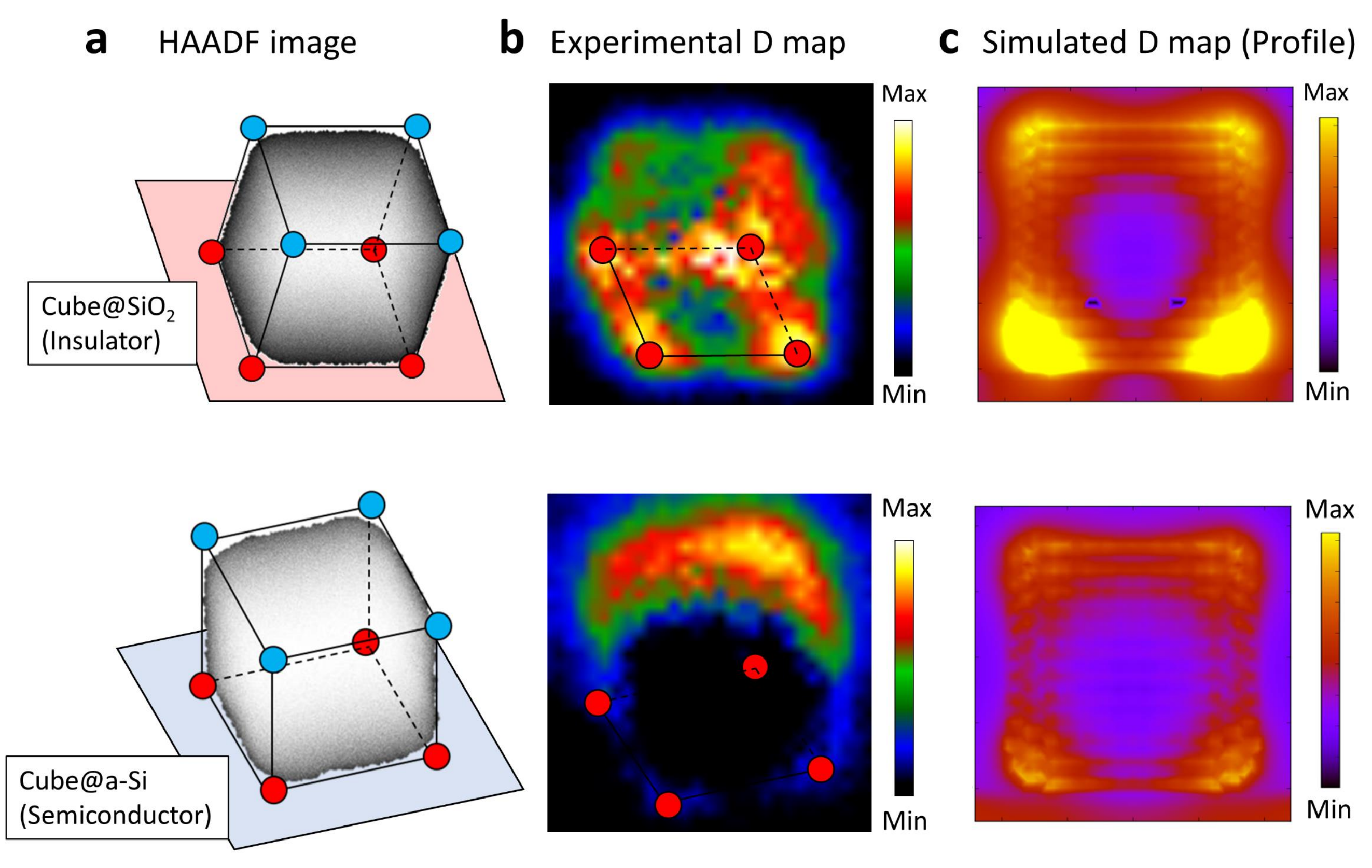}
\caption{Mapping energy transfer with STEM/EELS. (a) HAADF images of two tilted Ag nanocube@substrate systems. The red and blue dots represent near-substrate and near-vacuum corners, respectively. (b) Experimental D-mode EELS maps. The cube@SiO2 (insulator) system exhibits typical substrate-localization of the D mode, meaning that no energy transfer occurs in the system. By contrast, the near-zero EEL probability in the D mode map at the cube@a-Si (semiconductor) interfaces is a signature of energy transfer to the substrate. (c) Simulated EEL D mode maps of the cube@insulator and cube@semiconductor systems agree with the experiments. Reproduced with permission from Reference \cite{Li_energy_transfer_2015}. Copyright 2015 American Chemical Society.}
\label{Li}
\end{figure}

In a recent work by Li et al., mode maps proved to be an ideal way to image the spatial profile of energy transfer between a MNP and a semiconductor \cite{Li_energy_transfer_2015}. Using 3-D STEM/EELS to examine the Ag nanocube@substrate system \cite{Sherry_silver_nanocubes_2005, Zhang_Fano_resonances_2011, Ringe_nanocubes_2010}, they performed experiments on nearly-identical nanocubes on three different substrates: one insulating (SiO$_2$) and two semiconducting; boron phosphide (BP) and amorphous silicon (a-Si). \textbf{Figure \ref{Li}} displays the results for two of the three systems systems considered. They found that the cube@SiO2 system exhibits typical top/bottom  plasmon localization expected of substrate-dressed LSPs, in agreement with previous studies of cube@insulator systems. Specifically, the corner D mode is substrate-localized while its quadrupole-like counterpart (Q mode) exhibits vacuum localization. The two cube@semiconductor systems show almost zero EEL probability at the proximal corners, in sharp contrast to the D mode of the cube@SiO2 system. From the energetic point of view, this effect is observed as linewidth broadening in the EELS spectrum, and was attributed to plasmonic energy transfer from the nanocubes to the semiconducting substrates. This work extends the use of LSP mode mapping beyond the usual mode characterization to the study of energy transfer, allowing researchers to probe not only the amount of energy transfer occurring in the system but also \textit{where} energy transfer occurs with nanoscale resolution.

\section{Quantum Effects}
Quantum effects in plasmonics can broadly separated into two categories: effects arising from the quantum nature of photons and effects arising from the quantum properties of electrons. The latter category are strongly dependent on nanoparticle size and shape through the electron wave function and are often called quantum size effects \cite{kreibig1985optical,tame2013quantum}. Included in this category are electron tunneling, spill-out effects, and plasmon-induced hot electron generation, many of which have been successfully probed using an electron beam \cite{Ouyang_Quantum_size_1992,Scholl_Quantum_plasmon_2012,Scholl_quantum_tunneling_2013,wang2006quantum}. Meanwhile, in the first category, effects are often well understood by describing the LSP as a photon trapped in a lossy, dielectric cavity \cite{tame2013quantum,thakkar2015quantum}. Experimental studies of these effects have considered plasmonic particles excited by quantum states of light and have shown that quantum coherences are retained in photon-plasmon-photon conversion processes \cite{fakonas2014two,fujii2014direct,di2014observation,cai2014high}. Research on these effects is proceeding rapidly \cite{otten2015entanglement} but electron-beam based studies have only just begun \cite{bendana2010single}.

Quantum size effects have received interest due to the potential use of nanoparticles in quantum optoelectronic devices and biomedical applications, where small sizes are necessary in order to integrate plasmonics into cells  \cite{Cognet_protein_detection_2003, Hu_Gold_nanostructrues_2006}. Historically these effects were first studied by Kawabata and Kubo \cite{kawabata1966electronic}, whose linear response theory was extended by Rechsteiner and Smithard \cite{ganiere1975size} to predict a blueshift in the absorption spectrum of fine spherical particles (under $10$ nm in size). This prediction was in contrast to the expected result from classical Mie theory, which predicts a decreasing mode energy as particle size decreases. Most theoretical studies since Kawabata and Kubo have qualitatively reproduced this result from different considerations (including modern DFT approaches) \cite{kreibig1985optical,Scholl_Quantum_plasmon_2012}, but a few have not \cite{wood1982quantum,ekardt1984work}. Conclusive experimental observation one way or the other remained ellusive for many years, mostly due to the limitations of the far-field probes and the decreasing intensity of scattering and absorption signals of small particles.

\begin{figure}
\centering
\includegraphics[scale=0.4]{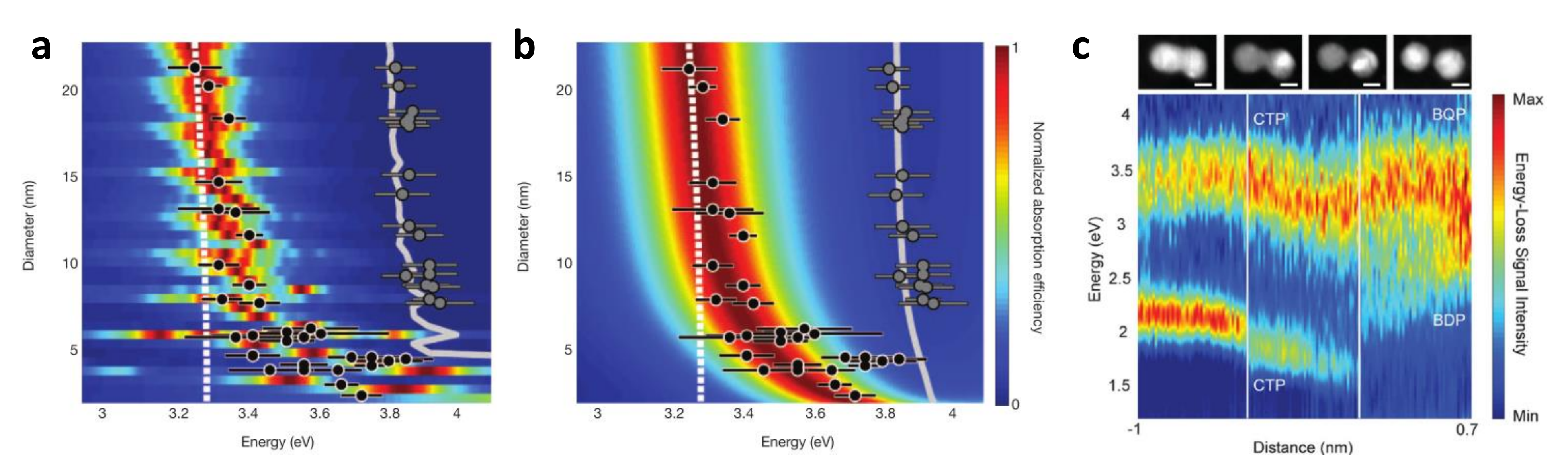}
\caption{Experimentally determined LSP energies (black dots) overlaid on the absorption spectra generated from the (a) analytic and (b) DFT-derived model of the permittivity.  Experimental bulk plasmon energies (grey dots) and theoretically predicted values (grey line) are also included. Results from classical theory are indicated with a dashed white line. The experimental results significantly deviate from the classical models for particles smaller than $\sim$10 nm in size, indicating the presence of the quantum size effects. Reproduced with permission from Reference \cite{Scholl_Quantum_plasmon_2012}. Copyright 2012 Nature Publishing. (c) EEL spectra acquired with the electron beam placed at one end of a 9-nm-diameter silver homodimer. The electron beam exerts a force on the system, thereby reducing the gap between the MNPs. When the gap is $\sim$0.5 nm, we see the signature of an electron tunneling event in the EEL spectrum. Reproduced with permission from Reference \cite{Scholl_quantum_tunneling_2013}. Copyright 2013 American Chemical Society.}
\label{Dionne_Quantum}
\end{figure}

Electron energy-loss experiments have since settled the debate. In 1992, Ouyang et al. became the first to study small silver particles (hemispheres, in this case) with EELS and report a blue shift in LSP energy as particle diameter dropped below 10 nm all the way to 4 nm \cite{Ouyang_Quantum_size_1992}. Later, Dionne and co-workers studied small silver spheres ranging from 1.7 nm to 20 nm \cite{Scholl_Quantum_plasmon_2012}. This experiment also conclusively shows deviations from classical physics by reporting a blue shift from 3.3 eV to 3.8 eV ({\bf Figure \ref{Dionne_Quantum} (a) and (b)}). Interestingly, they report that the bulk plasmon energy also experiences a 0.1 eV blueshift when the particle is smaller than 6 nm in size. Dionne et al. went further to compare their experiment to semiclassical theory and DFT calculations, and both theoretical approaches agreed excellently with the experiment. The main results are shown in Figure \ref{Dionne_Quantum}, which compares the experiment to both theories. 
In 2013, Dionne et al. applied these same methods to study to quantum-sized dimer systems and measured the signature of a tunneling event between Ag particles \cite{Scholl_quantum_tunneling_2013} in the EELS spectrum (Figure \ref{Dionne_Quantum}(c)). They found that electric-field enhancements resulting from the hybridized bright mode would saturate when the inter-particle spacing decreases to 0.5 nm. At these separation distances, they argue that a charge transfer LSP mode is formed due to quantum tunneling, resulting in distinct features in the EELS spectrum. In 2014, Nijhuis and co-workers demonstrated the control over the quantum tunneling effect by bridging two Ag nanocubes with a self-assembled molecular monolayer \cite{tan2014quantum}. By changing the molecular monolayer from saturated, aliphatic 1,2-ethanedithiolates (EDT) to aromatic 1,4-benzenedithiolates (BDT) while measuring EELS spectra, they found that the energy of the charge transfer mode shifted from about 1.0 to 0.6 eV. Also in 2013, an EELS study by Li and colleagues reported on electron tunneling in Au nanoprism dimer systems \cite{Wu_Fowler_Nordheim_2013}.
 
\section{Beyond EELS: Cathodoluminescence and Electron Energy-Gain Spectroscopy}
Other forms of electron spectroscopy offering additional and complementary information to EELS have been realized in the STEM and scanning electron microscope (SEM). In the past several years, new measurement techniques based on cathodoluminescence (CL) \cite{vesseur2007direct,van2006direct,kuttge2009local,kaz2013bright,myroshnychenko2012plasmon,losquin2015unveiling} and electron energy-gain spectroscopy (EEGS) \cite{barwick2009photon,park2010photon,piazza2015simultaneous} have provided a deeper understanding of nanoscale optical processes involving surface plasmons.

\begin{figure}
\centering
\includegraphics[scale=0.45]{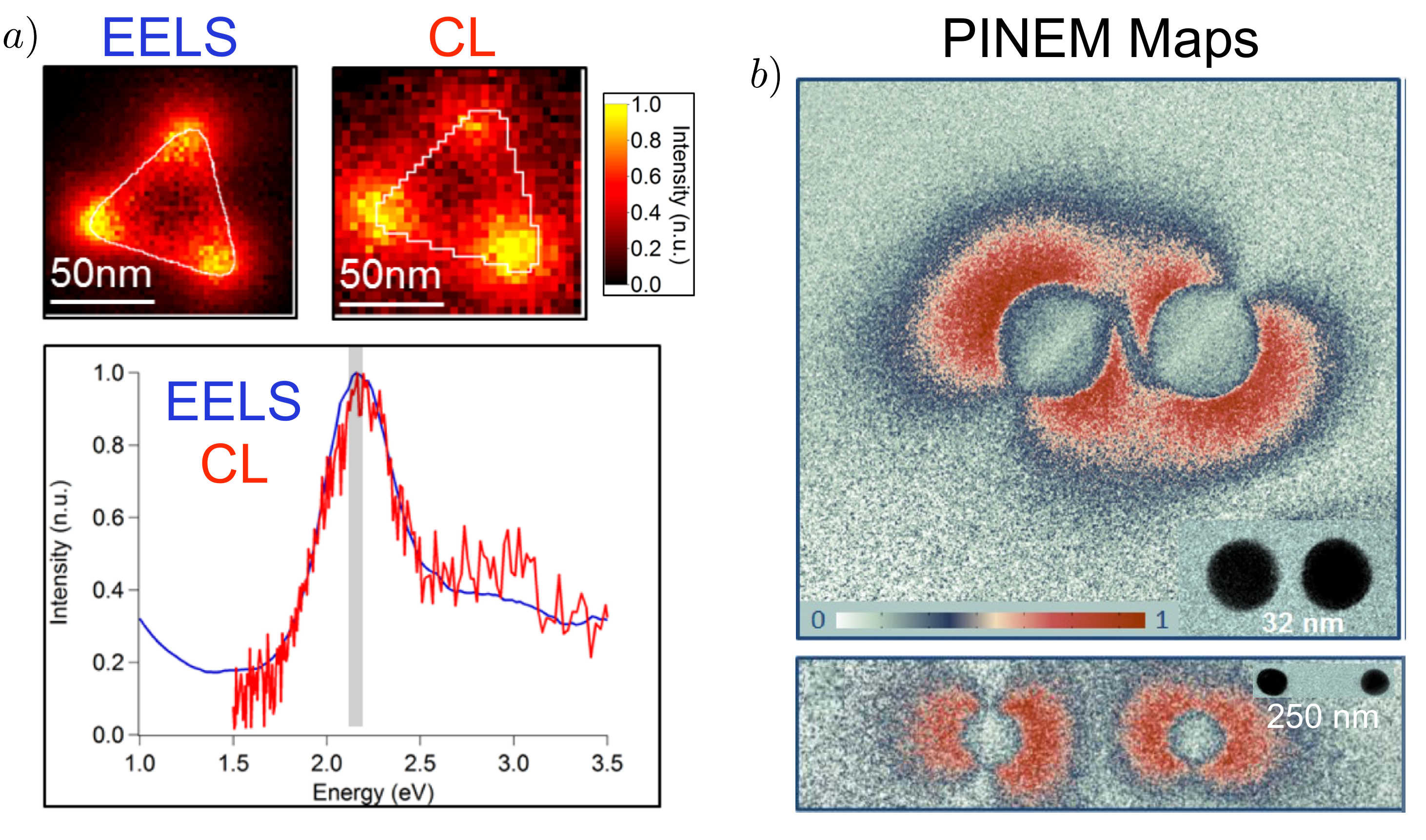}
\caption{Comparing EELS, CL, and PINEM. (a) In a correlated study of Au nanoprisms, CL and EELS were applied to the same nanoparticle, revealing the complementary information contained in each. Kociak and co-workers have demonstrated that the EEL and CL maps are nearly identical when evaluated at the same LSP energy. Reproduced with permission from Reference \cite{losquin2015unveiling}. Copyright 2015 American Chemical Society. (b) Zewail and co-workers use PINEM, an electron energy gain experiment with pulsed laser source, to demonstrate the emergence of interference phenomena as two silver nanoparticles are brought into close proximity. PINEM allows maps to be generated at fixed polarization, preserving a signature of the nodal structure of the plasmonic near-field. Reproduced with permission from Reference \cite{yurtsever2012entangled}. Copyright 2012 American Chemical Society.}
\label{CL}
\end{figure}

CL is the light emission from a material excited by electron impact. In particular, the electron beam can be used to excite LSPs in MNP systems, which then radiate to the far-field if they are bright. This radiation can be focused onto an imaging detector by a parabolic mirror, and its intensity as a function of frequency can be organized into a CL spectrum. When this spectrum is normalized to the intensity of the incident electron beam, it is a measure of the system's probability to emit a photon of a particular color at a particular beam position. As the electron beam is rastered across the target, CL photons can be collected at each spatial point and organized into a CL map analogous to the EEL mode maps discussed previously. The CL and EEL maps are complementary; they give similar spatial information but yield added information about the mode's parity as only bright modes will emit photons. Since EELS probes all modes, CL is especially well-suited for determining LSP parities in complex nanoparticle aggregates. {\bf Figure \ref{CL} (a)} shows a correlated study of LSPs in triangular Au nanoprims using EELS and CL \cite{losquin2015unveiling}.

The roles of the electron beam and photon as pump and probe can be reversed. This is the central concept in EEGS, where the MNP system is pumped by light at a particular frequency and then probed by the electron beam. In this setup, the passing electron can gain energy as a result of interaction with the excited LSP, and this energy gain can be plotted as a function of excitation frequency to yield an EEG spectrum. In contrast to EELS, the incident light in EEGS can be used to target a specific LSP mode, yielding spectra with energy resolution determined only by the linewidth of the incident light, which is often lower than what is achievable in most STEMs. Since the light source can be polarized, rastering the electron-beam to produce an EEG map revealing the nodal structure of the LSP mode's electric-near field. Recently, Zewail and co-workers \cite{barwick2009photon,park2010photon, yurtsever2012entangled} demonstrated that EEGS can be initiated with optical pulses, allowing them to measure LSP properties as a function of space (beam position) and delay time between the optical pulse and electron probe. This new technique, called photon-induced near-field electron microscopy (PINEM), offers picosecond time resolution in addition to the sub-nanometer spatial resolution familiar to EELS. An example of using PINEM to interrogate the coupling between two neighboring nanoparticles is shown in {\bf Figure \ref{CL} (b)}.  

By adding the radiation field's energetic contributions to Eq.~ (\ref{Hamiltonian}), both CL and EEGS can be modeled with the theory presented in Section 2. Given this new Hamiltonian, which incorporates plasmonic coupling to photons, the CL and EEG spectra can be thought of as second-order processes and calculated with perturbation theory. Garc\'{i}a de Abajo and co-workers \cite{asenjo2013plasmon} calculate the EELS, CL, and EEGS probability densities for the dipole ($\ell=1$) LSP of a sphere. They find
\begin{eqnarray}
\Gamma_{{\tiny\textrm{EELS}}}(\omega) &=& \frac{4 e^2 \omega^2}{\pi\hbar v^4 \gamma^2}\left[K_1^2\left(\left|\frac{\omega R_0}{v\gamma}\right|\right)+ \frac{1}{\gamma^2}K_0^2\left(\left|\frac{\omega R_0)}{v \gamma}\right|\right)\right]\textrm{Im}\left\{\tilde{\alpha}(\omega)\right\} \\
\Gamma_{{\tiny\textrm{CL}}} (\omega) &=& \frac{8 e^2 \omega^5}{3\pi\hbar c^3 v^4 \gamma^2}\left[K_1^2\left(\left|\frac{\omega R_0}{v\gamma}\right|\right) + \frac{1}{\gamma^2}K_0^2\left(\left|\frac{\omega R_0}{v\gamma}\right|\right)\right] |\tilde{\alpha}(\omega)|^2\\
\Gamma_{\tiny\textrm{EEGS}} (\omega) &=& \frac{8\pi e^2 \omega^2}{\hbar^2 v^4 c \gamma^2}I_0 K_1^2\left(\left|\frac{\omega R_0}{v \gamma}\right|\right)|\tilde{\alpha}(\omega)|^2 \delta(\omega -\omega_i),
\end{eqnarray}
where $\gamma = 1/\sqrt{1-v^2/c^2}$ accounts for relativistic corrections and $I_0$ is the intensity of the driving laser field. Comparison of the three rates shows that although each has a different frequency dependent prefactor, all three inherit their peak structure from the complex poles of the MNP polarizability $\tilde{\alpha}(\omega)$. Thus, although there are relative shifts in the spectra from each of the three experiments, they measure the same plasmon resonances inherent to a particular nanostructure.  

Indeed, in a recent paper by Kociak and co-workers \cite{losquin2015unveiling}, generalizations of these analytic results for arbitrary geometries \cite{boudarham2012modal} are used as a means to discuss the similarities and differences between CL and EELS. They find that CL and EELS are the near-field analogs to optical scattering and absorption respectively, and they emphasize that the peaks in EEL and CL spectra will not align due to differing dependence on frequency and damping. They also conclude that both $\Gamma_{\tiny\textrm{CL}}$ and $\Gamma_{\tiny\textrm{EELS}}$ have the same complex pole structure, and therefore measure the same plasmon resonances of a given MNP system. 

\section{Conclusion}
In this review we have discussed the historical origins of EELS and plasmons, we have presented an intuitive approach to the theory of surface plasmons and their interaction with energetic electrons, and have we have highlighted cutting-edge experiments published over the past five years that go beyond the simple characterization of LSP modes. The future of STEM/EELS as a window into the nanoscopic world is especially promising, and we expect continued advances in the molecular, optical, materials, information, and energy sciences as a result.

EELS benefits from the highly localized field of a relativistic electron, which drastically relaxes selection rules while also driving the sample with a broad range of frequencies. When EELS is performed in a scanning transmission electron microscope, it is capable of reaching sub-nanometer spatial resolution and up to $\sim$10 meV spectral resolution. Enormous improvements in electron monochromators and spectrometers and powerful numerical tools to simulate electron-plasmon interactions have placed STEM/EELS at the forefront of nanoscience.

%Disclosure
\section*{DISCLOSURE STATEMENT}
The authors are not aware of any affiliations, memberships, funding, or financial holdings that might be perceived as 
affecting the objectivity of this review.

% Acknowledgement
\section*{ACKNOWLEDGMENTS}
This work was supported by the U.S. National Science Foundation's CAREER program under award number CHE- 1253775 and through XSEDE resources under award number PHY-130045 (D.J.M.) and the NSF Graduate Research Fellowship Program under award number DGE-1256082 (N.T.). This work was also supported by the U.S. Department of Energy, Basic Energy Sciences under award number DE-SC0010536 (J.P.C., G.L.). G.L. was supported by a Center for Sustainable Energy at Notre Dame postdoctoral fellowship. C.C. would like to thank Professor D. H. Dunlap from the University of New Mexico for his review of the manuscript and insightful comments. C.C. would also like to thank Mr. Nicholas Montoni for aiding in locating certain references and Dr. Robert Cook for his feedback on the theory section. D.J.M. would like to thank Dr. Mathieu Kociak from the Laboratoire de Physique des Solides for a careful reading of the manuscript. 
% References
\bibliographystyle{ieeetr} 
\bibliography{ARPC.bib}
\end{document}